\documentclass{PoS}
\usepackage{siunitx}
\usepackage{wrapfig}
\usepackage{csquotes}
\usepackage{calrsfs}
\usepackage{booktabs}
\usepackage{caption}

\title{Search for Astrophysical Tau Neutrinos with an Improved
 Double Pulse Method}

\ShortTitle{Search for Astrophysical Tau Neutrinos}

\author{
The IceCube Collaboration\footnote{For collaboration list, see PoS(ICRC2019) 1177.}\\
{\itshape \href{http://icecube.wisc.edu/collaboration/authors/icrc19_icecube}{http://icecube.wisc.edu/collaboration/authors/icrc19\_icecube}}\\
E-mail: \email{mmeier@icecube.wisc.edu, jsoedingrekso@icecube.wisc.edu}
}

\abstract{

Tau neutrino identification with the IceCube experiment would open new windows to neutrino physics as well as enable novel searches for cosmic neutrino sources. This work aims at a identification of tau neutrinos with astrophysical origin at energies above $\sim{\SI{100}{\tera\electronvolt}}$. For identification, we search for a double pulse structure in the signal of one of IceCubes Digital Optical Modules originating from a tau neutrino interaction and a subsequent tau decay within the detector. In this work, we present constraints on the tau neutrino flux based on an event sample with a livetime of about \SI{7.5}{years} of IceCube data. \\

\vspace{4mm}
{\bfseries Corresponding authors:}
Maximilian Meier$^{1}$, \speaker{Jan Soedingrekso}$^{\, 1}$\\
{$^{1}$ \itshape TU Dortmund University}\\

}

\FullConference{36th International Cosmic Ray Conference -ICRC2019-\\
		July 24th - August 1st, 2019\\
		Madison, WI, U.S.A.}

\begin{document}

\section{Introduction} \label{sec:intro}

The IceCube collaboration first observed an astrophysical flux of high-energy neutrinos in 2013~\cite{hese_science}. Although the flavor ratio of neutrinos produced by pion decays would be $\nu_e : \nu_{\mu} : \nu_{\tau} = 1 : 2 : 0$ at a source, due to neutrino oscillations over astrophysical baselines the flavor ratio at Earth is modified to $\nu_e : \nu_{\mu} : \nu_{\tau} \simeq 1 : 1 : 1$. Tau neutrino production in atmospheric air showers is negligible~\cite{ers}, which makes the detection of tau neutrinos another important verification of the astrophysical origin of the observed high-energy neutrinos. The tau neutrino component of the astrophysical flux has yet to be identified. Previous searches have only produced upper limits on the astrophysical tau neutrino flux~\cite{donglian_paper, marcel_icrc} but, recently, the first tau neutrino candidates were presented~\cite{juliana_vlvnt}.

IceCube is a neutrino detector deployed deep in the ice at the geographic south pole instrumenting a volume of \SI{1}{\cubic\kilo\meter}~\cite{instrumentation}. IceCube consists of digital optical modules (DOMs) that observe Cherenkov light produced by secondary particles from neutrino interactions. This analysis aims to identify tau neutrinos via the so called double pulse signature~\cite{doug_taus_in_icecube}, where a single DOM records two distinct light depositions: the first from the hadronic cascade induced by a charged current (CC) $\nu_{\tau}$ interaction and the second one from the subsequent decay of the tau lepton (into an electron or into hadrons). This signature has two different backgrounds: single cascades produced by $\nu_e$ CC and all flavor neutral current (NC) interactions and tracks created by atmospheric muons and $\nu_\mu$ CC interactions (and $\nu_{\tau}$ CC interactions followed by a tau decaying into a muon). Section~\ref{sec:evt_sel} introduces an event selection removing single cascade events by identifying double pulse waveforms and track events by their topological differences to cascades using machine learning.

Based on the $\nu_{\tau}$ dominated event sample the astrophysical $\nu_{\tau}$ flux normalization is measured in a binned Poisson likelihood fit, described in section~\ref{sec:method}. In this analysis, \SI{7.5}{years} of IceCube data\footnote{All of the analyzed data was recorded while the detector has been in full operation with 86 strings.} recorded between May 2011 and December 2018 (with an effective livetime of \SI{2666.8}{days}) are analyzed. Two double pulse events found by this analysis are characterized in section~\ref{sec:results}.

\section{Event Selection} \label{sec:evt_sel}

\begin{figure}
    \centering
    \includegraphics[width=.6\textwidth]{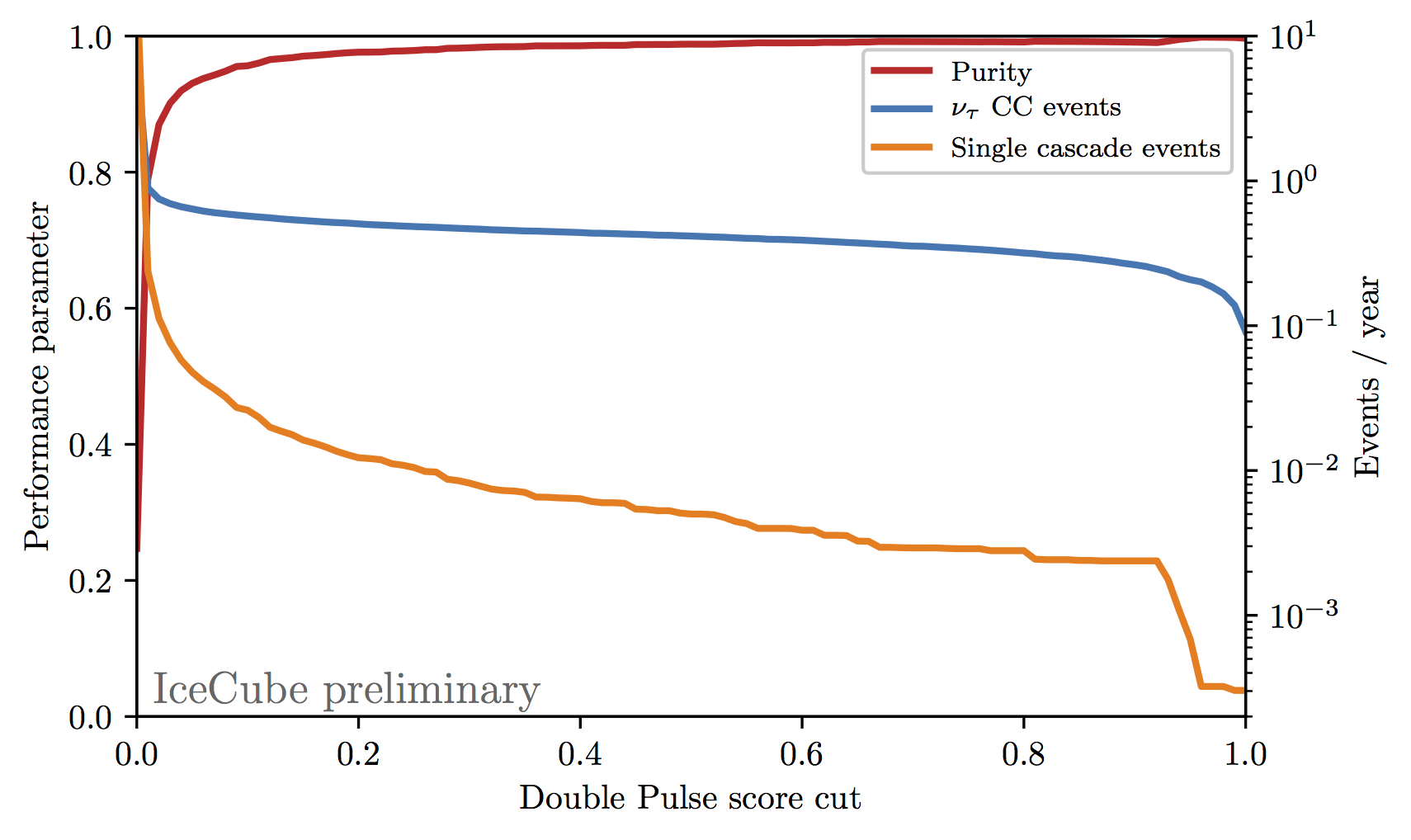}
    \caption{Influence of the Double Pulse score cut on expected events per year and the purity achieved for $\nu_{\tau} \, \mathrm{CC}$ with respect to all cascade-like events. \label{fig:dp_sel}}
\end{figure}

The event selection presented here improves the expected tau neutrino event rate of the three year double pulse search~\cite{donglian_paper} by a factor of \num{2.5}. The event selection is structured in the following way. First, IceCube waveforms are analyzed to select events containing at least one double pulse waveform. Double pulses are identified by training a Random Forest (RF)~\cite{rf} with Monte Carlo simulations for signal (double pulse waveforms from $\nu_\tau$ CC interactions) and cascade background (single cascade waveforms from $\nu_e$ CC and all-flavor $\nu$ NC interactions). For the RF the observables from~\cite{donglian_paper} are used to characterize the waveforms by detecting rising and trailing edges based on the waveform derivative. Additional observables are used, which are described in~\cite{dp_icrc2017}. These observables are summary statistics (e.g.~the mean) of the waveform, the number of local maxima, the difference between unsmoothed and smoothed waveforms, and the compatibility of the waveform with a fit to an exponential function starting from the first local maximum.  The RF assigns a score value to each waveform called \enquote{Double Pulse score}, where a value of \num{1} represents a signal-like waveform. The performance of the RF is illustrated in Fig.~\ref{fig:dp_sel}. It shows the remaining $\nu_\tau$ CC events and cascade-like events containing at least one double pulse waveform as a function of the Double Pulse score cut, which is the threshold at which a waveform is accepted as a double pulse waveform. Here, a threshold of \num{0.2} is chosen, reducing the background from single cascades by more than three orders of magnitude to a subdominant level. 

After the process of double pulse identification, the track-like background events are addressed. The sought after signal of $\nu_\tau$ CC events is still dominated by track-like events originating from atmospheric muons and $\nu_\mu$ CC interactions. The reason for that is that a muon can produce double pulses by a combination of Cherenkov light emission and high energy stochastic energy losses close to a DOM. In case of a $\nu_\mu$ CC interaction the hadronic cascade can also be responsible for one of the pulses. 

Atmospheric muons from cosmic ray air showers are primarily described with CORSIKA simulations~\cite{corsika}. For this analysis the most difficult background component is single muons depositing most of their energy in a single energy loss near the edge of the detector. To simulate this background more efficiently we rely on a parametrization of the single muon yield from CORSIKA, which is introduced in~\cite{mese}. The separation is, again, carried out by a RF using observables focusing on the event topology and the location of the event in the detector.

\begin{figure}
    \centering
    \includegraphics[width=0.55\textwidth]{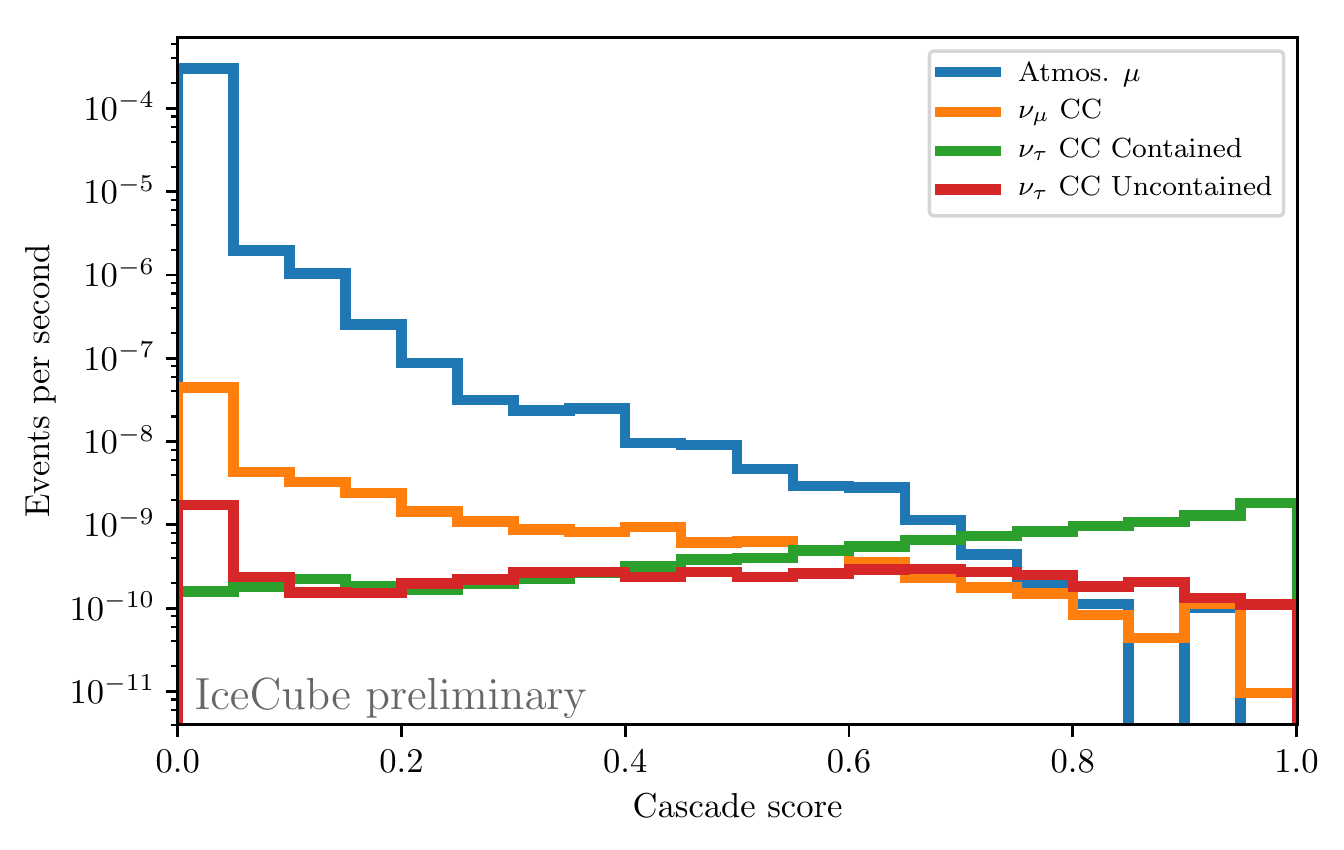}
    \includegraphics[width=0.42\textwidth]{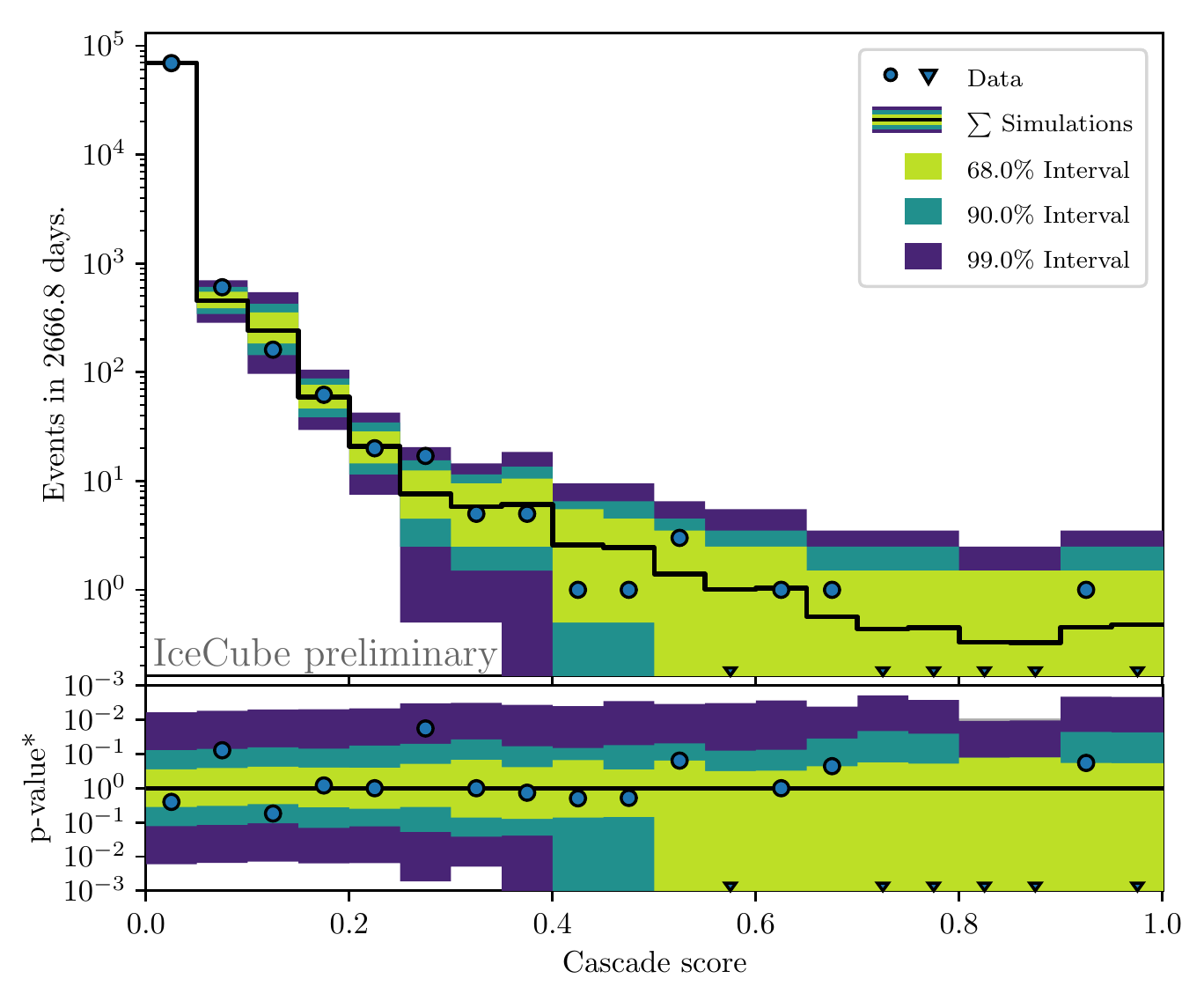}
    \caption{Cascade score distribution for different simulated components (left) and a comparison between the sum of simulations and \SI{7.5}{years} of IceCube data with an effective livetime of \SI{2666.8}{days} (right). \label{fig:cscd_sel}}
\end{figure}

The left panel of Fig.~\ref{fig:cscd_sel} shows the distribution of the \enquote{Cascade score} for different simulated components. The signal $\nu_\tau$ CC interactions are split up into a contained and an uncontained fraction, as only the contained events are used as signal events in the training. Besides that no explicit containment cut or veto is used. The right panel of Fig.~\ref{fig:cscd_sel} compares the sum of all simulated components with \num{7.5} years of data, which shows good agreement. The error bands on the expectation are a combination of Poisson uncertainties and uncertainties due to limited MC simulations~\cite{say_llh}\footnote{This way of comparing histograms of data and simulation is suggested in \cite{aggarwal}.}. The bottom panel shows the ratio between data and simulations. The \enquote{p-value*} is the p-value to observe the value found in data with respect to the expected value.

The threshold on the Cascade score was chosen by optimizing the model rejection factor~\cite{mrf} keeping all events with a score above \num{0.62} for further analysis. For an astrophysical per-flavor flux of $\SI{1.01e-18}{\per\giga\electronvolt \per\square\centi\meter \per\steradian \per\second} \left( E_{\nu} / \SI{100}{\tera\electronvolt} \right)^{-2.19}$~\cite{diffuse_numu_8yr}, the conventional atmospheric neutrino flux from~\cite{hkkms} and the prompt neutrino flux from~\cite{berss} the expected number of signal and background events is \num{2.10} and \num{0.98} respectively in the final sample for the analyzed time period of \num{7.5} years of data.

\section{Analysis Method} \label{sec:method}


The goal of this analysis is to constrain the tau neutrino flux normalization. This is done with a binned Poisson likelihood maximization with the bin-wise expectation $\mu_i$ defined as:
\begin{equation}
    \mu_i (\lambda ) = \mu_{B, i} + \lambda \mu_{S, i},
\end{equation}
with the background expectation $\mu_B$, the signal expectation $\mu_S$ and $\lambda$ as the only free parameter scaling the tau neutrino flux normalization. The astrophysical spectral index is kept fixed during the analysis as the number of overall expected events is small ($\sim \num{3}$) and thus not sufficient to measure multiple parameters. Instead, three spectral indices are tested ($E^{-2.19}$~\cite{diffuse_numu_8yr}, $E^{-2.50}$~\cite{global_fit} and $E^{-2.89}$~\cite{hese7.5}). The final sample also contains a considerable fraction of background events, so favoring observables with strong separation power between signal and background is beneficial. 

\begin{wrapfigure}{r}{0.49\textwidth}
  \vspace{-2em}
  \begin{center}
    \includegraphics[width=0.48\textwidth]{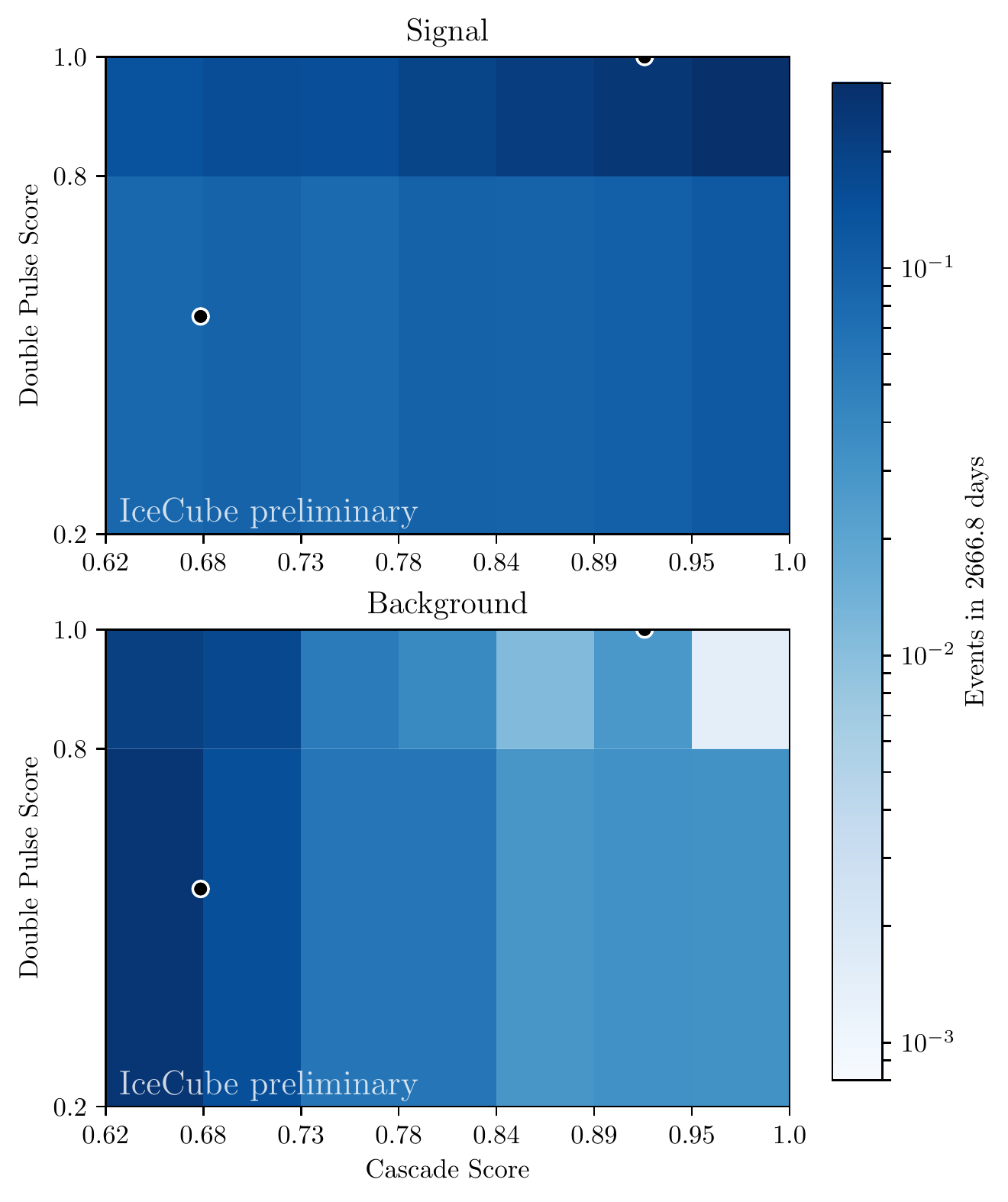}
  \end{center}
  \caption{Binned 2D-distribution of the Double Pulse score and the Cascade score. The top panel shows the signal component and the bottom panel the sum of all background components. The depicted distributions assume an astrophysical $E^{-2.19}$ spectrum. The two observed events are indicated by black dots.
  Event\#1 is located in the signal dominated region at (\num{0.92}, \num{1.0}) and Event\#2 in the background dominated region at (\num{0.675}, \num{0.565}).
  \label{fig:obs_dist}}
  \vspace{-3em}
\end{wrapfigure}

The observables used are the Double Pulse score and the Cascade score, that were already introduced earlier. One event can have multiple double pulse waveforms, therefore only the highest value per event for the Double Pulse score is used. Their 2D-distributions are presented in Fig.~\ref{fig:obs_dist}. The binning is chosen to have two irregular bins in Double Pulse score and seven equidistant bins in Cascade score, where the total number of bins is mainly restricted by limited simulations. The confidence interval for the likelihood fit is constructed via hypothesis test inversion~\cite{casella_berger}
\begin{equation}
    C(x) = \{ \lambda_0 \in \Lambda : -\Delta \ln(\mathcal{L}(\lambda_0 | x)) \geq k_{\alpha} \}
\end{equation}
where $x$ denotes the observed data sample, $\Lambda$ the allowed parameter space for $\lambda$, $-\Delta \ln(\mathcal{L}(\lambda_0 | x))$ the value of the likelihood ratio test statistic for the null hypothesis $\lambda = \lambda_0$ given the observation $x$ and $k_{\alpha}$ the critical value for the  likelihood ratio test at confidence level (CL) 1-$\alpha$.

Based on the same observable distributions observed event candidates are assigned a p-value to describe their compatibility with originating from the background distribution. This is constructed from the likelihood
\begin{equation}
    \mathcal{L}(\lambda|i) = \frac{P_{B, i} + \lambda P_{S, i}}{1 + \lambda},
\end{equation}
with the probability distributions $P_B$ and $P_S$ describing the simulated events in the described binned 2D observable space. The parameter $\lambda$ is bound to the interval $[0, 1]$ and is used to fit the \enquote{signalness} in the corresponding observable bin $i$. The background distribution $P_B$ is confirmed to be consistent with data in a Cascade score region close to the signal region.

\section{Results} \label{sec:results}

\begin{figure}
    \centering
    \includegraphics[width=.45\textwidth]{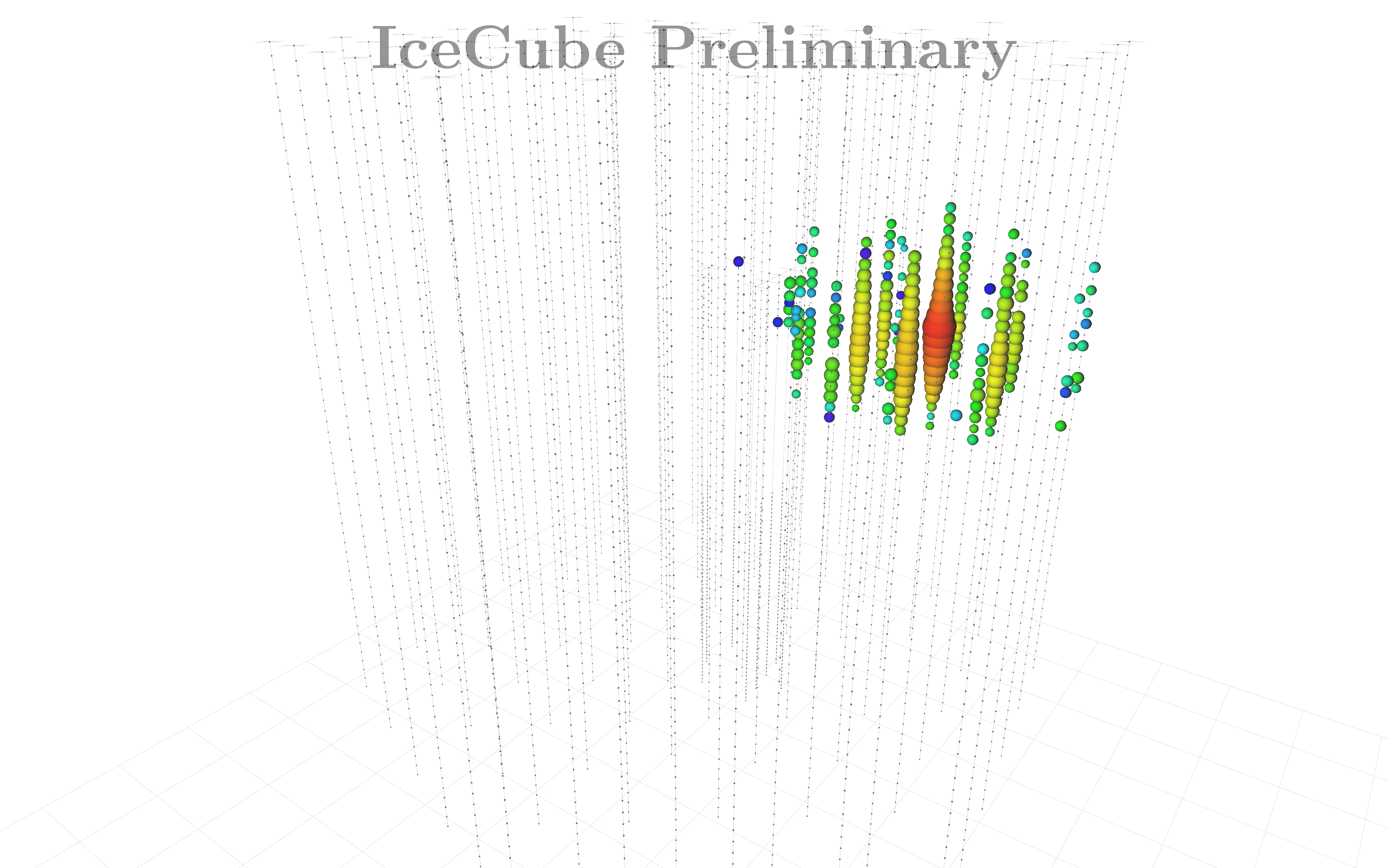}
    \includegraphics[width=.5\textwidth]{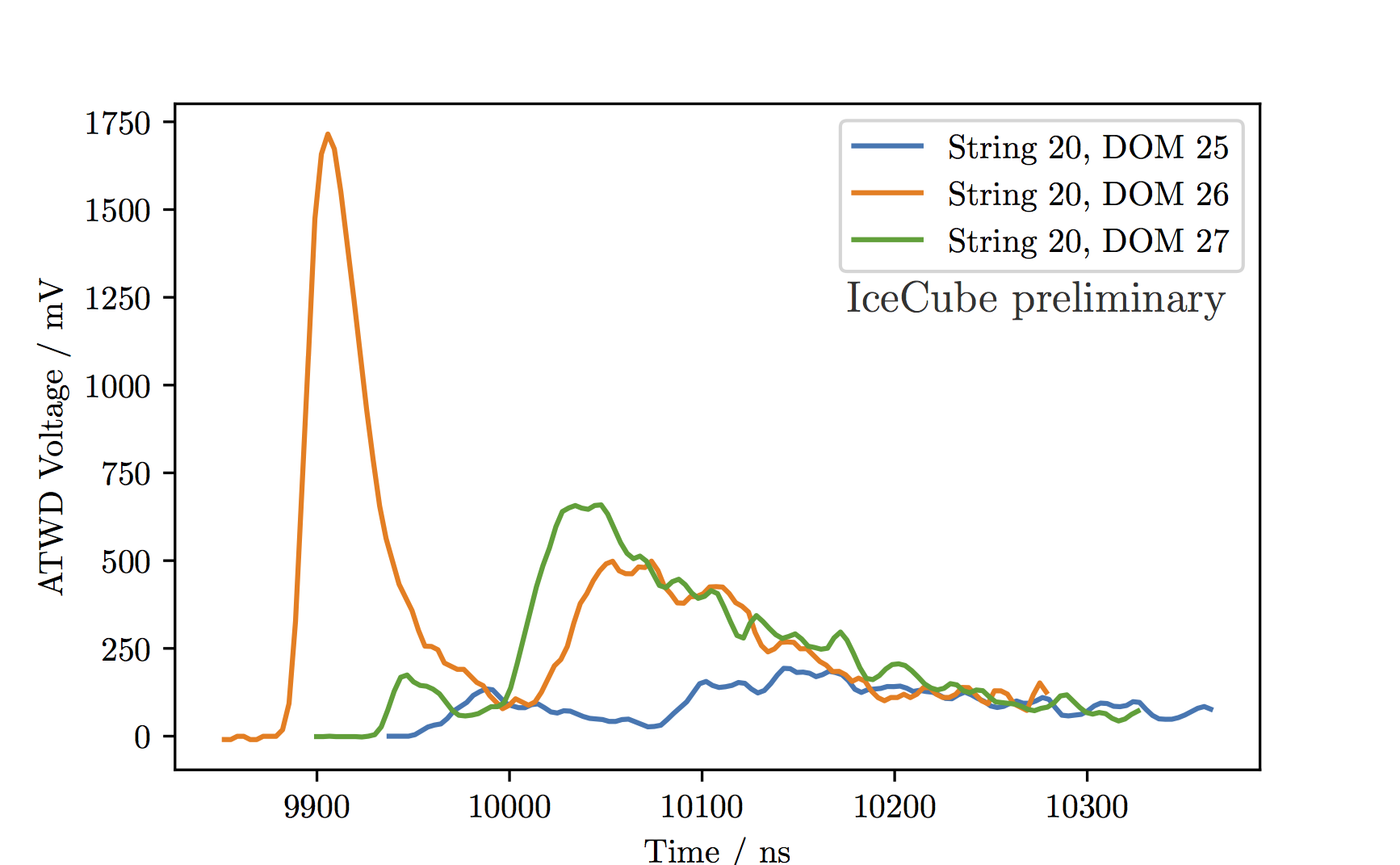}
    \caption{Left: Event view of the double pulse event recorded in the 2014 season. Right: Double pulse waveforms recorded for this event. The corresponding double pulse scores are \num{0.45} (blue), \num{1.0} (orange) and \num{0.81} (green). This event was also observed by~\cite{donglian_logan_icrc}. The orange and green waveforms are seen by that analysis as well. The blue waveform is only classified as a double pulse waveform in this analysis. \label{fig:event1}}
\end{figure}

IceCube data recorded between May 2011 and December 2018 was analyzed and two tau neutrino candidates were observed. Event\#1, observed in the 2014 season\footnote{A south pole season starts in spring and ends the spring of the following year.}, is presented in the left panel of Fig.~\ref{fig:event1}\footnote{IceCube event views represent observed charge as a sphere around the DOM observing it. The size of the sphere is proportional to the deposited charge and the arrival time is encoded with different colors from early (red) to late (blue).}. The right panel shows the three observed double pulses, located on three adjacent DOMs, with double pulse scores between \num{0.45} and \num{1.0}. The Cascade score for this event is found to be \num{0.92}. The resulting p-value is $p \simeq \num{0.035}$, almost independent of the assumed astrophysical spectrum. This event is also observed as a double cascade tau neutrino candidate by~\cite{juliana_icrc} and as a double pulse tau neutrino candidate by~\cite{donglian_logan_icrc}.

\begin{figure}
    \centering
    \includegraphics[width=.45\textwidth]{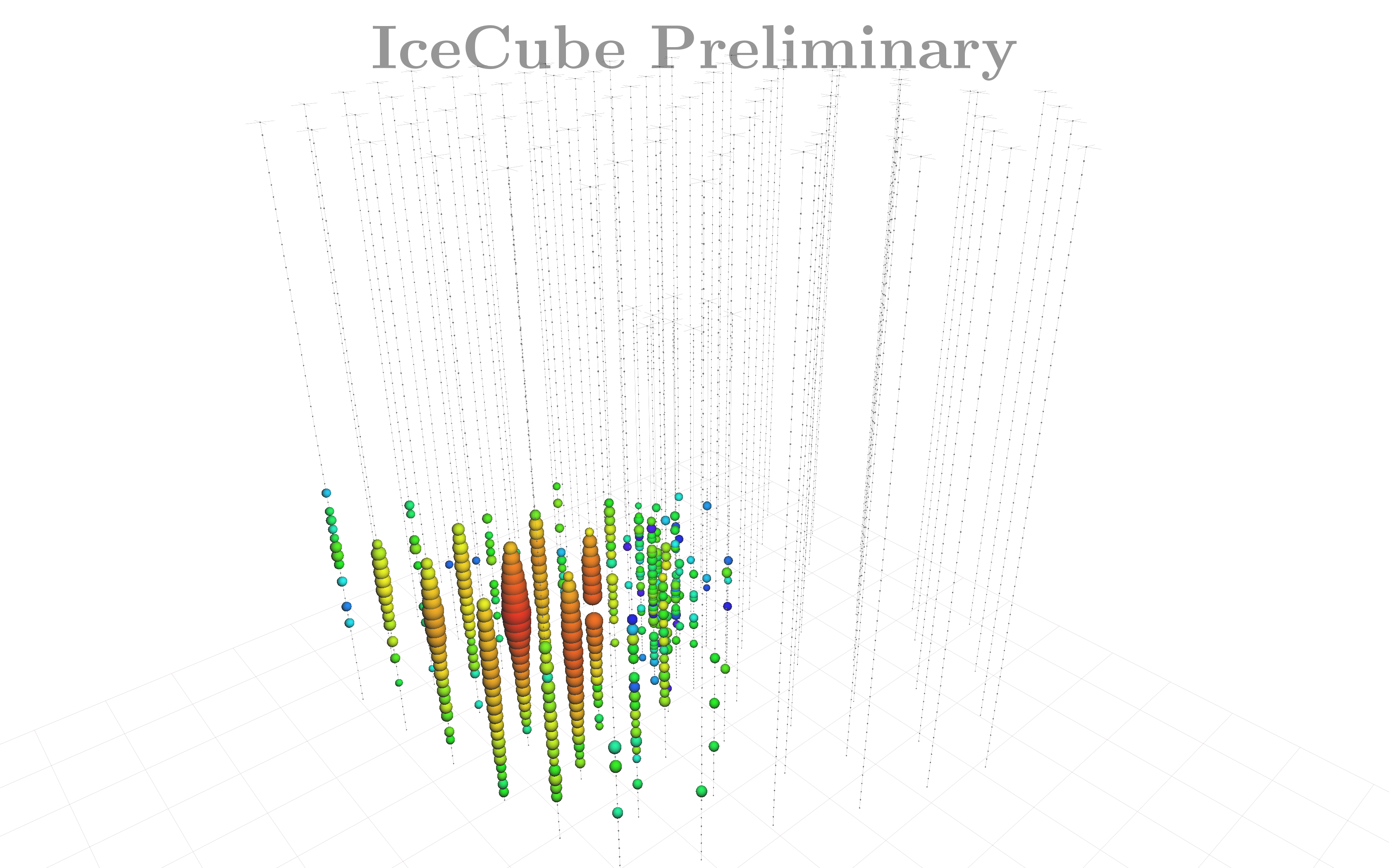}
    \includegraphics[width=.5\textwidth]{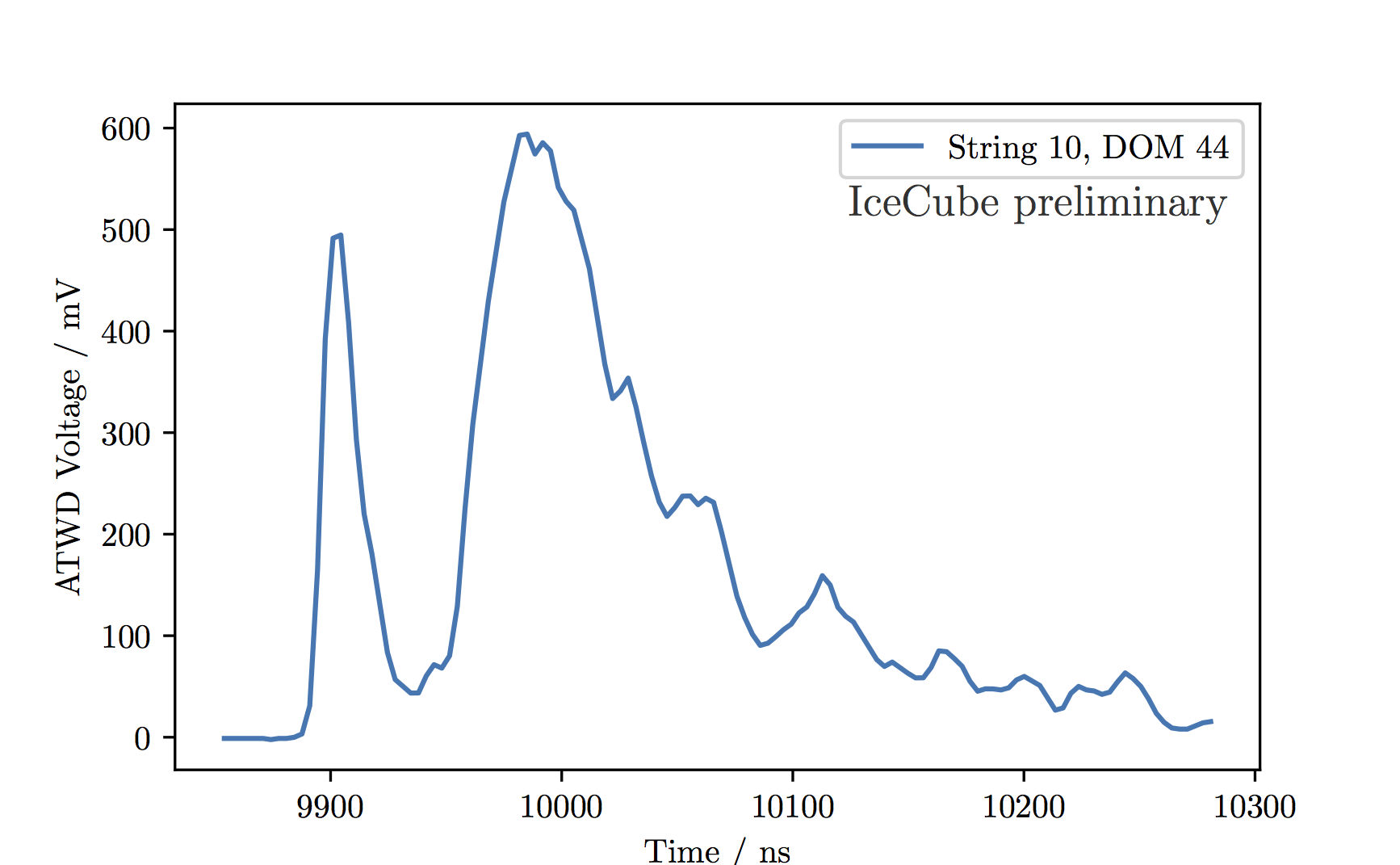}
    \caption{Left: Event view of the double pulse event recorded in the 2015 season. Right: Double pulse waveform recorded for this event with a double pulse score of \num{0.565}. \label{fig:event2}}
\end{figure}

Event\#2 was recorded in the 2015 season and is depicted in the left panel of Fig.~\ref{fig:event2}. The right panel shows the observed double pulse waveform with a double pulse score of \num{0.565}. The Cascade score for this event is \num{0.675}. The resulting p-value is $p = \num{1.0}$ for all tested astrophysical spectra. The cascade score for both events is indicated in Fig.~\ref{fig:cscd_sel} (right) and Fig.~\ref{fig:obs_dist}.

\begin{figure}
\begin{minipage}[b]{0.58\textwidth}
    \centering
    \includegraphics[width=1.\textwidth]{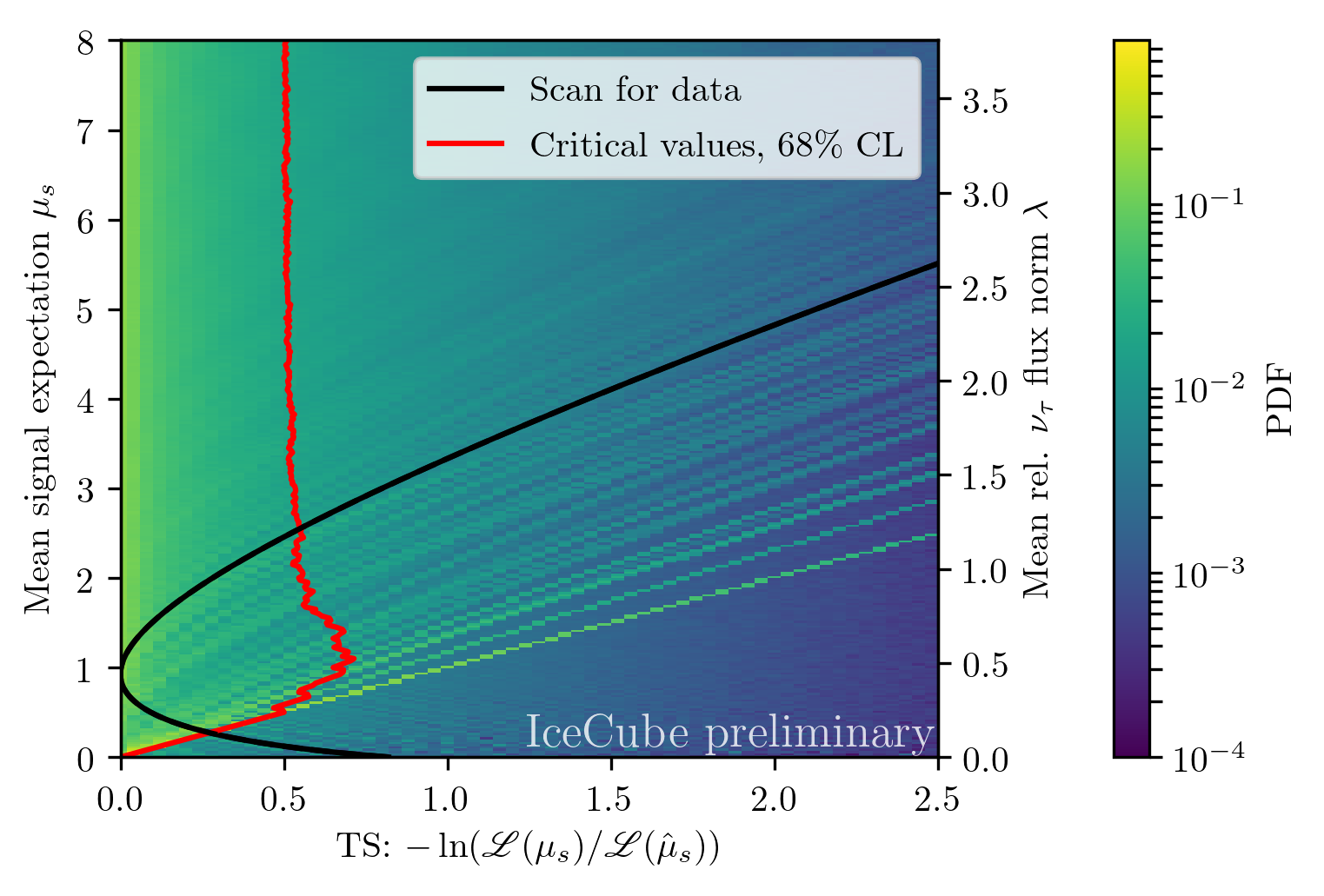}
    \captionsetup{width=.9\linewidth}
    \caption{Likelihood scan for the observed data. The red line shows critical values at \SI{68}{\percent} CL obtained from the underlying TS distributions for each value of $\mu_s$. \label{fig:fit_result}}
\end{minipage}
\begin{minipage}[b]{0.40\textwidth}
    \centering
	\renewcommand{\arraystretch}{1.3}
	\begin{tabular}[hbt]{cc}
		\toprule
		$\gamma$ & $\Phi_{0, \nu_{\tau}}$ at \SI{100}{\tera\electronvolt} / \\
		& ($10^{-18} \, \si{\per\giga\electronvolt\per\centi\meter\squared\per\second\per\steradian}$) \\
		\midrule
		2.19 & $\num{0.45}^{+\num{0.79}}_{-\num{0.31}}$ \\
		2.50 & $\num{0.83}^{+\num{1.46}}_{-\num{0.59}}$ \\
		2.89 & $\num{1.62}^{+\num{2.78}}_{-\num{1.11}}$ \\
		\bottomrule
	\end{tabular}
	\vspace{3em}
	\captionsetup{width=.9\linewidth}
	\captionof{table}{Measurement of the astrophysical tau neutrino flux normalization for different astrophysical spectra. \label{tab:fit_results}}
	\renewcommand{\arraystretch}{1.0}
\end{minipage}
\end{figure}

The resulting data sample was used to measure the tau neutrino flux normalization with a binned Poisson likelihood fit as described in the previous section for three different assumptions on the shape of the astrophysical spectrum: $E^{-2.19}$, $E^{-2.50}$ and $E^{-2.89}$. Fig.~\ref{fig:fit_result} shows the likelihood scan when assuming an astrophysical spectral index of $E^{-2.19}$ as a black line. The color scale in the background indicates the TS distributions obtained by pseudo-experiments for different injected values for the mean signal expectation $\mu_s$. The red line depicts the critical values to reject a certain value of $\mu_s$ at \SI{68}{\percent} CL. Then the confidence interval is constructed by finding the set of $\mu_s$ values that can not be rejected at a certain CL.

The measured tau normalizations are presented in Tab.~\ref{tab:fit_results}. Independent of the spectral index the fit prefers a non-zero normalization. They are smaller than the baseline normalizations due to one observed event being a small, yet unsignificant underfluctuation relative to the expectation. Within the 1$\sigma$ uncertainty they agree with a flavor composition of $\nu_e : \nu_\mu : \nu_\tau = 1 : 1 : 1$. In each case, the observation is incompatible with a tau normalization of $0$ at the $\sim \num{1.9}\sigma$ level. As already mentioned, systematic uncertainties are not included in this measurement. Their influence is small compared to the statistical uncertainty due the small data sample, and they will be discussed in more detail in a forthcoming publication.


\begin{figure}
    \centering
    \includegraphics[width=.8\textwidth]{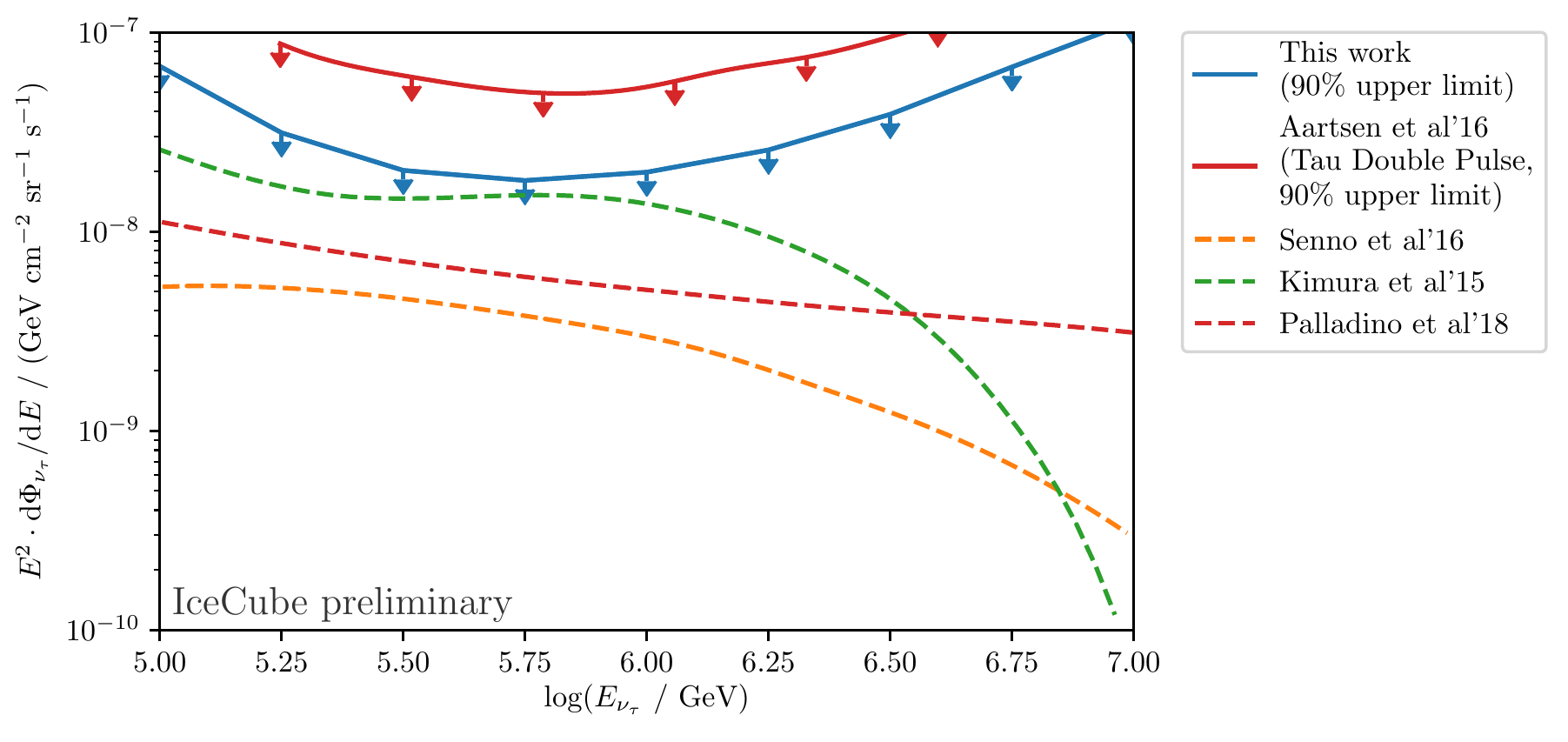}
    \caption{Differential upper limit on the tau neutrino flux between 100 TeV and 10 PeV. The red solid line shows the differential upper limit observed by a previous search for astrophysical tau neutrinos conducted on three years of IceCube data~\cite{donglian_paper}. As a comparison, models for per-flavor astrophysical fluxes from choked jets and low luminosity GRBs~\cite{senno_choked_jets} (orange), low luminosity AGNs~\cite{kimura_ll_agn} (green) and from a multi-component model~\cite{palladino_multicomp} (red) are shown. \label{fig:differential_limit}}
\end{figure}

Besides the model dependent constraints, based on the findings a differential upper limit was constructed to obtain a less model dependent constraint. The differential upper limit is constructed the same way as described in~\cite{ehe_9yr} and~\cite{donglian_paper}. The same likelihood formalism as described in section~\ref{sec:method} is used, but for each tested energy $E_c$ a signal flux proportional to $E^{-1}$ centered around $E_c$ with a width of one decade is injected and the resulting \SI{90}{\percent} upper limits are calculated.
The differential upper limit is presented in Fig.~\ref{fig:differential_limit} and compared to the differential limit of the previous \SI{3}{year} double pulse analysis~\cite{donglian_paper} and to three models predicting the astrophysical neutrino flux from chocked jets and low luminosity gamma-ray bursts~\cite{senno_choked_jets}, low luminosity AGNs~\cite{kimura_ll_agn} and a multicomponent model~\cite{palladino_multicomp}.

\section{Summary}

We have performed a new measurement of the astrophysical tau neutrino flux based on \SI{7.5}{years} of IceCube data. An improved method using machine learning to identify tau neutrinos via their double pulse signature is presented. Two tau neutrino double pulse candidates were found. One candidate, also observed in the IceCube double cascade analysis~\cite{juliana_icrc}, is found to be signal-like with a p-value of $p \simeq \num{0.035}$ based on the observables used in this analysis. A model dependent measurement of the astrophysical tau neutrino flux normalization is presented in Tab.~\ref{tab:fit_results}, finding a non-zero normalization with a statistical significance on the $\num{1.9}\sigma$ level. The observed values are compatible with a flavor ratio of $\nu_e : \nu_\mu : \nu_\tau = 1 : 1 : 1$ and are not in conflict with previously reported upper limits from IceCube. Albeit two tau neutrino candidates were detected the differential upper limit on the tau neutrino flux, previously reported in~\cite{donglian_paper}, was also improved.

\bibliographystyle{ICRC}
\bibliography{references}

\providecommand{\href}[2]{#2}\begingroup\raggedright\begin{thebibliography}{10}

\bibitem{hese_science}
{\bf IceCube} Collaboration, M.~G. Aartsen et~al., {\em Science} {\bf 342}
  (2013).

\bibitem{ers}
R.~Enberg, M.~H. Reno, and I.~Sarcevic, {\em Phys. Rev. D} {\bf 78} (Aug, 2008)
  043005.

\bibitem{donglian_paper}
{\bf IceCube} Collaboration, M.~G. Aartsen et~al., {\em Phys. Rev.} {\bf D93}
  (2016) 022001.

\bibitem{marcel_icrc}
{\bf IceCube} Collaboration,  \pos{PoS(ICRC2017)974} (2018).

\bibitem{juliana_vlvnt}
{\bf IceCube} Collaboration, M.~G. Aartsen et~al., {\em EPJ Web Conf.} {\bf
  207} (2019) 02005.

\bibitem{instrumentation}
{\bf IceCube} Collaboration, M.~G. Aartsen et~al., {\em JINST} {\bf 12} (2017)
  P03012.

\bibitem{doug_taus_in_icecube}
D.~Cowen, {\em Journal of Physics: Conference Series} {\bf 60} (3, 2007)
  227--230.

\bibitem{rf}
L.~Breiman, {\em Machine Learning} {\bf 45} (2001) 5--32.

\bibitem{dp_icrc2017}
{\bf IceCube} Collaboration,  \pos{PoS(ICRC2017)1009} (2018).

\bibitem{corsika}
D.~{Heck} et~al., {\em {CORSIKA: a Monte Carlo code to simulate extensive air
  showers.}}
\newblock Feb., 1998.

\bibitem{mese}
{\bf IceCube} Collaboration, M.~G. Aartsen et~al., {\em Phys. Rev.} {\bf D91}
  (2015) 022001.

\bibitem{say_llh}
C.~A. Arg{\"u}elles, A.~Schneider, and T.~Yuan,
  \href{http://arxiv.org/abs/1901.04645}{{\tt arXiv:1901.04645}}.

\bibitem{aggarwal}
R.~Aggarwal and A.~Caldwell, {\em Eur. Phys. J. Plus} {\bf 127} (2012) 24.

\bibitem{mrf}
G.~C. Hill and K.~Rawlins, {\em Astropart. Phys.} {\bf 19} (2003) 393--402.

\bibitem{diffuse_numu_8yr}
{\bf IceCube} Collaboration,  \pos{PoS(ICRC2017)1005} (2018).

\bibitem{hkkms}
M.~Honda et~al., {\em Phys. Rev. D} {\bf 75} (Feb, 2007) 043006.

\bibitem{berss}
A.~Bhattacharya et~al., {\em Journal of High Energy Physics} {\bf 2015} (Jun,
  2015) 110.

\bibitem{global_fit}
{\bf IceCube} Collaboration, M.~G. Aartsen et~al., {\em Astrophys. J.} {\bf
  809} (2015) 98.

\bibitem{hese7.5}
{\bf IceCube} Collaboration,  \pos{PoS(ICRC2019)1004} (these proceedings).

\bibitem{casella_berger}
G.~Casella and R.~L. Berger, {\em Statistical Inference}.
\newblock 2nd. Duxbury, 2002.

\bibitem{juliana_icrc}
{\bf IceCube} Collaboration,  \pos{PoS(ICRC2019)1015} (these proceedings).

\bibitem{donglian_logan_icrc}
{\bf IceCube} Collaboration,  \pos{PoS(ICRC2019)1036} (these proceedings).

\bibitem{senno_choked_jets}
N.~Senno, K.~Murase, and P.~Meszaros, {\em Phys. Rev.} {\bf D93} (2016) 083003.

\bibitem{kimura_ll_agn}
S.~S. Kimura, K.~Murase, and K.~Toma, {\em Astrophys. J.} {\bf 806} (2015) 159.

\bibitem{palladino_multicomp}
A.~Palladino and W.~Winter, {\em Astron. Astrophys.} {\bf 615} (2018) A168.

\bibitem{ehe_9yr}
{\bf IceCube} Collaboration, M.~G. Aartsen et~al., {\em Phys. Rev.} {\bf D98}
  (2018) 062003.

\end{thebibliography}\endgroup

%

\end{document}